\documentclass[num-refs]{wiley-article}

\papertype{Original Article}
\paperfield{Journal Section}

\usepackage{epstopdf}
\usepackage{natbib}
\usepackage[markup=underlined]{changes}
\usepackage{multirow}
\usepackage{subcaption}  
\AddToHook{cmd/added/before}{\def\Changes@AuthorColor{red}}
\AddToHook{cmd/deleted/before}{\def\Changes@AuthorColor{green}}
\AddToHook{cmd/replaced/before}{\def\Changes@AuthorColor{blue}}

\title{Scanner-based real-time automated volumetry reporting of the fetus, amniotic fluid, placenta and umbilical cord for fetal MRI at 0.55T}

\author[1,2]{Sara Neves Silva*}
\author[1,2]{Alena Uus*}
\author[1]{Hadi Waheed}
\author[1]{Simi Bansal}
\author[1]{Kamilah St Clair}
\author[1]{Wendy Norman}
\author[1,2,9]{Jordina Aviles Verdera}
\author[1]{Daniel Cromb}
\author[1]{Tomas Woodgate}
\author[1]{Milou van Poppel}
\author[1]{Johannes K Steinweg}
\author[1,2]{Jacqueline Matthew}
\author[1,2,5]{Kuberan Pushparajah}
\author[1,2,5]{David Lloyd}
\author[1]{Vanessa Kyriakopoulou}
\author[6]{Dimitris Siasakos}
\author[6]{Anna David}
\author[1,4]{Joseph V Hajnal}
\author[1,7,8]{Lisa Story}
\author[1]{Mary A Rutherford}
\author[1,2,9]{Jana Hutter}

\affil[*]{S.N.S. and A.U. contributed equally to this work.}
\affil[1]{Research Department of Early Life Imaging, School of Biomedical Engineering and Imaging Sciences, King’s College London, London, UK}
\affil[2]{Biomedical Computing Department, School of Biomedical Engineering and Imaging Sciences, King’s College London, London, UK}
\affil[4]{Research Department of Imaging Physics and Engineering, School of Biomedical Engineering and Imaging Sciences, King’s College London, London, UK}
\affil[5]{Department of Congenital Heart Disease, Evelina London Children’s Hospital, London, UK}
\affil[6]{Elizabeth Garrett Anderson Institute of Women's Health, University College London, London, UK}
\affil[7]{Department of Women and Children's Health, King's College London, London, UK}
\affil[8]{Fetal Medicine Unit, Guy’s and St Thomas’ NHS Foundation Trust, London, UK}
\affil[9]{Smart Imaging Lab, Radiological Institute, Friedrich-Alexander University Erlangen-Nuremberg, Germany}

\corraddress{Sara Neves Silva, King's College London, London, UK}
\corremail{sara.neves\char`_silva@kcl.ac.uk}

\runningauthor{Sara Neves Silva, Alena Uus et al.}

\begin{document}

\maketitle

Total Words: 3391 
\begin{abstract}

\noindent \textbf{Purpose:} 
This work aims to enable real-time automated intra-uterine volumetric reporting and fetal weight estimation for fetal MRI, deployed directly on the scanner. 
\\
\noindent \textbf{Methods:} 
A multi-region segmentation nnUNet was trained on 146 bSSFP images of 73 fetal subjects (coronal and axial orientations) for the parcellation of the fetal head, fetal body, placenta, amniotic fluid and umbilical cord from whole uterus bSSFP stacks. A reporting tool was then developed to integrate the segmentation outputs into an automated report, providing volumetric measurements, fetal weight estimations, and z-score visualisations. The complete pipeline was subsequently deployed on a 0.55T MRI scanner, enabling real-time inference and fully automated reporting in the duration of the acquisition. 
\\
\textbf{Results:} 
The segmentation pipeline was quantitatively and retrospectively evaluated on 36 stacks of 18 fetal subjects and demonstrated sufficient performance for all labels, with high scores ($>$0.98) for the fetus, placenta and amniotic fluid, and 0.91 for the umbilical cord. The prospective evaluation of the scanner deployment step was successfully performed on 50 cases, with the regional volumetric reports available directly on the scanner.
\\
\noindent \textbf{Conclusions:} 
This work demonstrated the feasibility of multi-regional intra-uterine segmentation, fetal weight estimation and automated reporting in real-time. This study provides a robust baseline solution for the integration of fully automated scanner-based measurements into fetal MRI reports.  
\\
\end{abstract}

\newpage
\section*{Introduction}

While antenatal imaging is routinely performed using ultrasound, fetal MRI is increasingly employed as a complementary modality for complex fetal anomalies \cite{Prayer2023}. In addition to offering superior soft tissue contrast, it provides full-view 3D spatial information, reduces operator dependency, and allows fetal assessment until late gestation. Biometric and volumetric measurements are thereby a crucial aspect of radiological reporting, particularly allowing quantitative assessment of fetal development and comparison to normal ranges, enabling the early detection of deviations.

In the third trimester, accurate assessment of fetal growth and the intrauterine environment is critical for optimal perinatal outcomes. Estimation of fetal weight is essential for identifying growth abnormalities, such as fetal growth restriction (FGR) and macrosomia, both linked to increased risks of stillbirth, operative delivery, and neonatal morbidity \cite{Aucott2004,WeissmannBrenner2012}. Fetal volume is additionally used to normalise organ volumes (e.g., lungs \cite{Story2020}) in cases with pathologies. Placental and amniotic fluid volumes are key indicators of fetal well-being. Small placental volume may suggest placental insufficiency and is associated with FGR and pre-eclampsia, while large placental size may indicate gestational diabetes or fetal anomalies \cite{GonzalezGonzalez2017,Soongsatitanon2019}. Similarly, abnormal amniotic fluid levels - oligohydramnios or polyhydramnios -may indicate placental dysfunction, fetal renal/gastrointestinal anomalies, or maternal-fetal transfer issues \cite{Morris2014}. Umbilical cord measurements, including insertion site, volume, coiling index, and length, provide additional information on fetal circulation and risk factors \cite{DeLaat2006}. However, they cannot be assessed using conventional clinical ultrasound. Short cords are associated with growth restriction and adverse outcomes, while excessively long cords increase the risk of entanglement or cord accidents \cite{Krakowiak2004}. Together, these metrics are important for assessing fetal health and planning delivery. They contribute to risk stratification and inform clinical decisions regarding timing and mode of birth, especially in high-risk pregnancies.

However, in routine clinical practice, ultrasound-based estimation of fetal weight \cite{Milner2018} and amniotic fluid volume \cite{Hughes2020} are typically derived indirectly from 2D biometry measurements, which can be inconsistent and inaccurate. Placental volume is not currently routinely clinically assessed. Furthermore, ultrasound accuracy in late gestation is reduced due to fetal head engagement in the pelvis and the lack of consideration of the contribution of fetal fat. It is further limited in cases of high maternal BMI, oligohydramnios, or unfavourable fetal positioning. In contrast, fetal MRI enables direct volumetric measurements based on true 3D information, and the measurements derived from segmentations are operator-independent, thus offering high reproducibility and accuracy in estimation of fetal weight \cite{KADJI2019428}. 3D segmentation of the fetus in MRI stacks has been used for fetal weight estimation \cite{Liao2019} and global organ volume normalisation \cite{Cannie2008,Story2020}. Although highly relevant \cite{Story2018,Li2024_jmri}, placenta and amniotic fluid volumes are not often measured with MRI. These large structures may be segmented in either balanced steady-state free-precession (bSSFP) or T2-weighted single-shot Turbo Spin Echo (ssTSE) sequences, using a large region-of-interest (ROI) coverage. Conventional clinical practice and research studies rely mainly on 2D slice-wise manual segmentations \cite{Liao2019,Story2020,Bouachba2025} for calculation of fetal body volumes, which is particularly time-consuming for larger fetuses at later gestation age (GA). 


Recently, several research works relying on UNet-style deep learning (DL) networks \cite{Ronneberger2015} proposed automated solutions for segmentation of large intrauterine structures including: fetus \cite{Lo2021,SpecktorFadida2024,Bischoff2025}, placenta \cite{Pietsch2021,Li2024} and amniotic fluid \cite{Costanzo2023,Csillag2023} in bSSFP, ssTSE and EPI fetal MRI stacks. Yet, combined segmentation of these structures is still lacking, which could be beneficial for combined relative volumetry analysis, as well as more consistent segmentation between ROI interfaces.

Furthermore, such frameworks were performed offline post-scan, foregoing the ability to react (e.g., by performing further dedicated sequences) to any measured deviations while the patient is still in the scanner. Recent studies have integrated real-time processing tools for fetal MRI scans, including quality control and re-acquisition of low-quality slices and stacks \citep{Gagoski2022,AvilesVerdera2025}, brain tracking \cite{NevesSilva2023}, 3D reconstruction \cite{Uus2025} and stack acquisition planning \cite{NevesSilva2024}. However, real-time volumetric analysis and reporting has not yet been performed.

\subsection*{Contributions}
In this work, we present the first scanner-based framework for interactive assessment in fetal MRI, by developing a real-time automated DL internal uterine parcellation and volumetry tool for the fetus, placenta, amniotic fluid and umbilical cord. The segmentation results are combined into an automated report with 3D visualisation, calculated centiles and estimated fetal weight (EFW) - all available during the scan. The method is implemented and tested for 0.55T late GA fetal MRI datasets and evaluated on a cohort of 50 prospective scans.

\section*{Methods}

\begin{figure*}[h!]
    \centering
    \includegraphics[width=.85\textwidth]{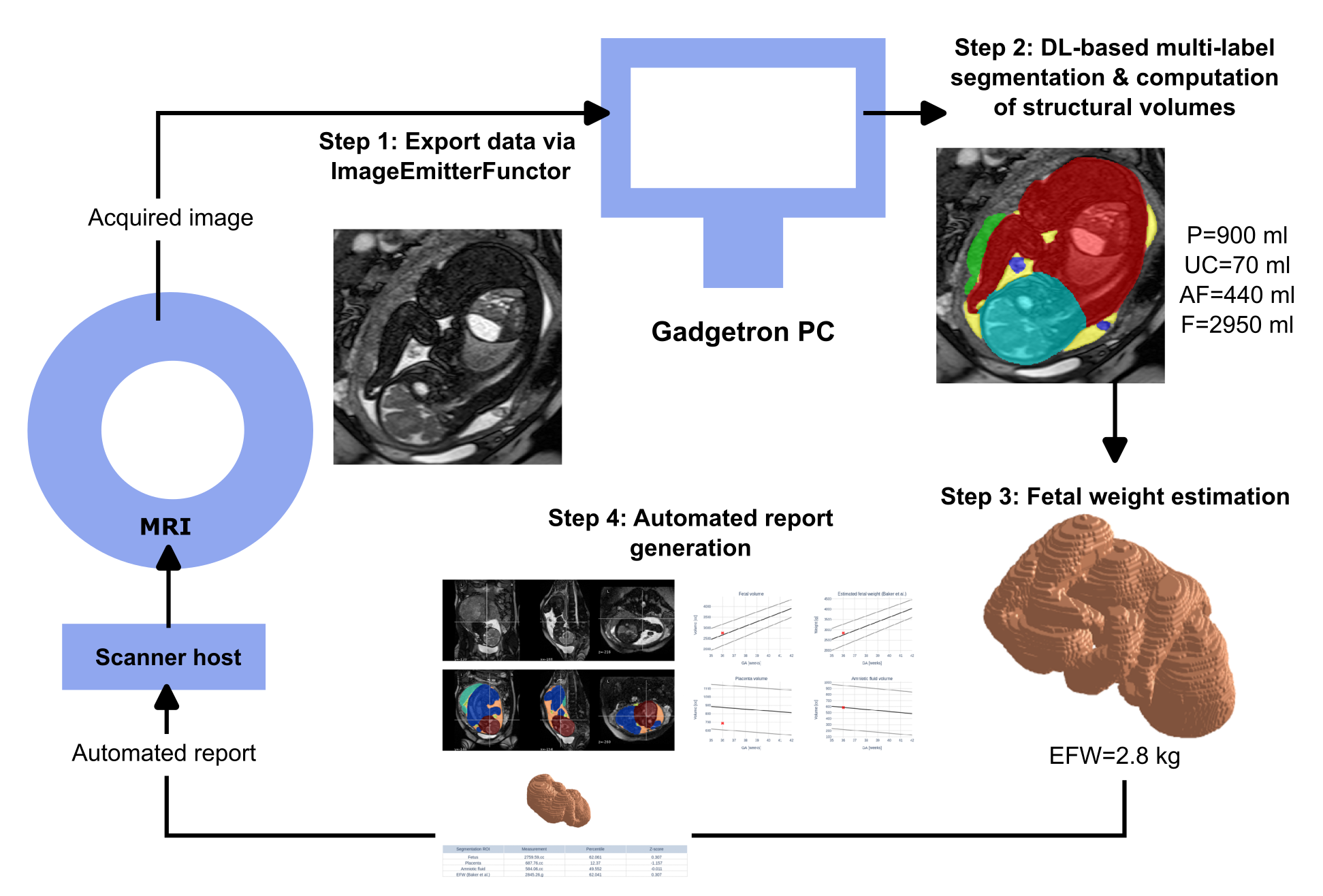}
    \caption{Overview of the automated volumetry reporting pipeline. Following the bSSFP acquisition, the data is immediately exported to an external Gadgetron PC using the FIRE framework (Step 1); then, DL-based segmentation is performed (Step 2), enabling fetal weight estimation (Step 3) and automated report generation (Step 4).}
    \label{fig:overview}
\end{figure*}

An overview of the proposed framework is presented in Fig.~\ref{fig:overview}. Immediately upon acquisition, a whole uterus bSSFP stack (maternal coronal orientation) is exported via FIRE \cite{chow2021prototyping} (Step 1) to an external Gadgetron PC. Next, DL-based segmentation is performed for 3D parcellation of the fetus, placenta, amniotic fluid and umbilical cord (Step 2). Based on the segmentation output, the volumetric measurements are performed, including fetal weight estimation (Step 3), calculated centiles, 3D models and growth charts, and all results are compiled into a .pdf report (Step 4), available for viewing on the scanner console.

\subsection*{Cohort, datasets and acquisition parameters}

The fetal MRI datasets used in this study were acquired at St. Thomas' Hospital, London, as part of the ethically approved MEERKAT [REC: 21/LO/0742], MiBirth [REC: 23/LO/0685] and NANO [REC: 22/YH/0210] studies. All experiments were performed in accordance with relevant guidelines and regulations. Informed written consent was obtained from all participants.

The fetal scans were performed on a 0.55T clinical MRI scanner (MAGNETOM Free.Max, Siemens Healthcare, Germany) in the supine position, using a 6-element blanket coil and a 9-element spine coil. A bSSFP sequence was acquired in coronal orientation to the whole-uterus ROI with TR=669.7 ms, TE=4.2 ms, FA=120 deg, GRAPPA acceleration factor 2, Partial Fourier 6/8, in-plane resolution=0.77x0.77 mm, slice thickness=2.4 mm, matrix size=544x544, and 64-80 slices per dataset. 

The gestational ages of the 50 prospective datasets ranged between 25.4 and 39.3 weeks, and primarily included late gestation cases between 36-40 weeks due to the specific aims of the main study ("MiBirth: MRI imaging at term for prediction of the mode of birth" , \url{https://www.mibirthstudy.com/}). The main inclusion criteria for all cases were: singleton pregnancy, no severe structural anomalies of the fetus, and acceptable image quality with clear visibility of the uterus and fetus and without extreme signal artefacts. 

\subsection*{Multi-regional internal uterine segmentation}

\begin{figure*}[h!]
    \centering
    \includegraphics[width=.85\textwidth]{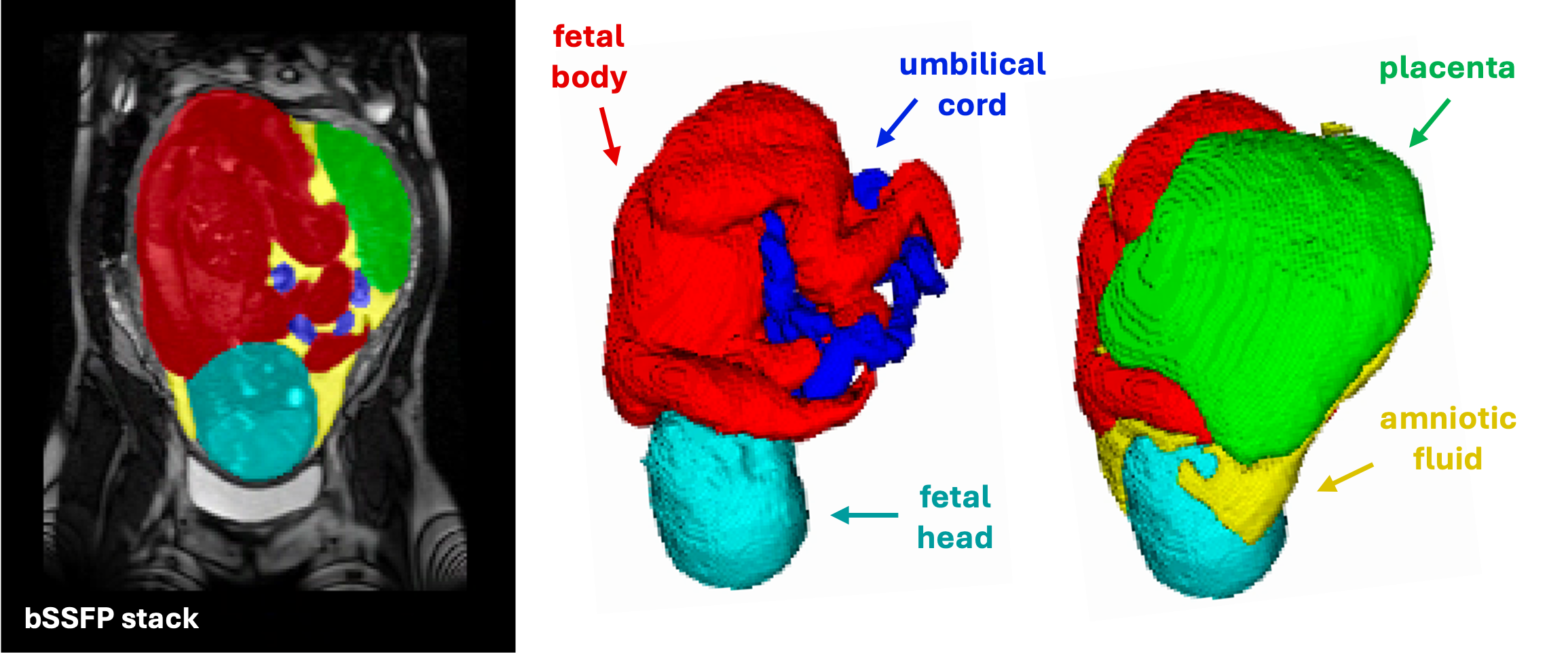}
    \caption{Proposed multi-regional intrauterine segmentation protocol for bSSFP stacks: fetal body (red), fetal head (cyan), placenta (green), umbilical cord (blue), amniotic fluid (yellow).}
    \label{fig:protocol}
\end{figure*}

The proposed parcellation protocol for the bSSFP whole uterus acquisitions, summarised in Fig.~\ref{fig:protocol}, was designed to include regions that are relevant to both volumetric analysis and 3D relative position assessment. It includes five labels in total: the fetus (subdivided into body and head labels), placenta, umbilical cord, and amniotic fluid. 

The nnUNet \cite{Isensee2021} framework was selected for the DL segmentation task. It was trained on 146 labelled bSSFP images, acquired in axial and coronal whole uterus orientations from 73 fetal subjects (GA 37-40 weeks). The ground-truth labels were created using outputs of in-house pre-trained DL networks for individual structures (developed for other projects within the department), followed by manual refinement (AU, SNS, SB, KSC, WN).  

A 3D U-Net architecture was implemented via the nnU-Net \cite{Isensee2021}, with pre-processing, architecture, and training parameters automatically configured based on the characteristics of the dataset. The model architecture integrated a 3D PlainConvUNet design with six encoder and five decoder stages, each composed of two convolutional layers. Most kernels were 3×3×3, except the first layer (1×3×3), with feature maps ranging from 32 up to 320. Pooling was asymmetrical, using six downsampling stages with kernel sizes: [1,1,1], [1,2,2], [2,2,2], [2,2,2], [2,2,2], and [1,2,2]. Images were resampled to 2.4x0.77x0.77 mm, and z-score normalisation was applied per case. The input patch size was 56x224x160 voxels. Training used stochastic gradient descent with Nesterov momentum (initial learning rate 0.01) and a combined Dice and cross-entropy loss. A batch size of 2 was chosen. From the initial 146 datasets, 117 cases (5-fold cross-validation) were used for training, and 29 cases for validation (1 fold). Standard nnU-Net data augmentations were applied, and model performance was assessed using the pseudo-Dice similarity coefficient across all segmented regions. The model was trained for 1000 epochs.

As a part of the retrospective quantitative evaluation of the pipeline, the model was subsequently tested on 36 images, 18 in each orientation. All 18 network predictions extracted from the coronal whole uterus images were refined by a fetal radiographer (R1, 1.5 years of experience), while a subset was additionally refined by R2 (3 years of experience) to measure inter-observer variability. The average volume differences were calculated for all labels and normalised to the manually refined volume.

\subsection*{Fetal weight estimation}
Similarly to previous studies performing whole fetus segmentation (e.g., \cite{SpecktorFadida2024}), EFW was computed from the total fetal label volume based on the classical model formulas derived by Baker et al. ($EFW_{Baker}(kg) = 1.031 \cdot V_{fetus} + 0.12$) \cite{Baker1994} and Kacem et al. ($EFW_{Kacem}(kg) = 0.989 \cdot V_{fetus}(L) + 0.147$) \cite{Kacem2013}.

\subsection*{Normative ranges for late gestation datasets}

In order to create normative growth centiles for automated reporting, the trained network was applied to segment an additional dataset comprising 90 coronal bSSFP stacks from MiBirth control subjects (35-39 weeks GA). The inclusion criteria were: no reported fetal and maternal anomalies. All DL segmentations were reviewed and manually refined when required. The extracted volumes for the fetus and placenta, along with the estimated fetal weight \cite{Baker1994,Kacem2013}, were then used to generate normative growth charts. The mean 50th, 5th and 95th centile models were created using classical linear fitting \cite{Royston1998}. 

\subsection*{Automated reporting}

The volumetry report was structured to include all information relevant for comprehensive clinical interpretation. It comprises: visualisation of the original bSSFP stack slices with the respective segmentations overlaid, visualisation of the 3D model of the fetus, a table presenting extracted volumes and EFW with computed centiles and z-scores, and corresponding plots derived from normative growth charts. The report is automatically generated in .pdf based on a Python script.

\subsection*{Scanner-based deployment}

The 0.55T scanner was connected to an external GPU-accelerated Gadgetron PC (NVIDIA GEFORCE RTX 2080 Ti, NVIDIA Corporate), and the prototype Siemens framework for image reconstruction environments (FIRE) \cite{chow2021prototyping} was used as the interface for real-time inline processing of the scanner-reconstructed images, as described above in Fig. \ref{fig:overview}. Each prospective dataset was streamed and stored in the workstation in real-time, and the online pipeline was configured using an XML file, stored in the scanner console. This file, linked with the respective sequence, specified the path to the Python processing script located on the workstation. Once the complete image matrix was collected, the processing pipeline was triggered automatically - the multi-label nnUNet segmentation model was immediately loaded and applied to the acquired image matrix, volumes were extracted for the segmented ROIs, the 3D model of the fetus and EFW were computed, and the report was generated.

\section*{Results}

\subsection*{Multi-regional internal uterine segmentation}

\begin{figure*}[h!]
    \centering
    \includegraphics[width=.85\textwidth]{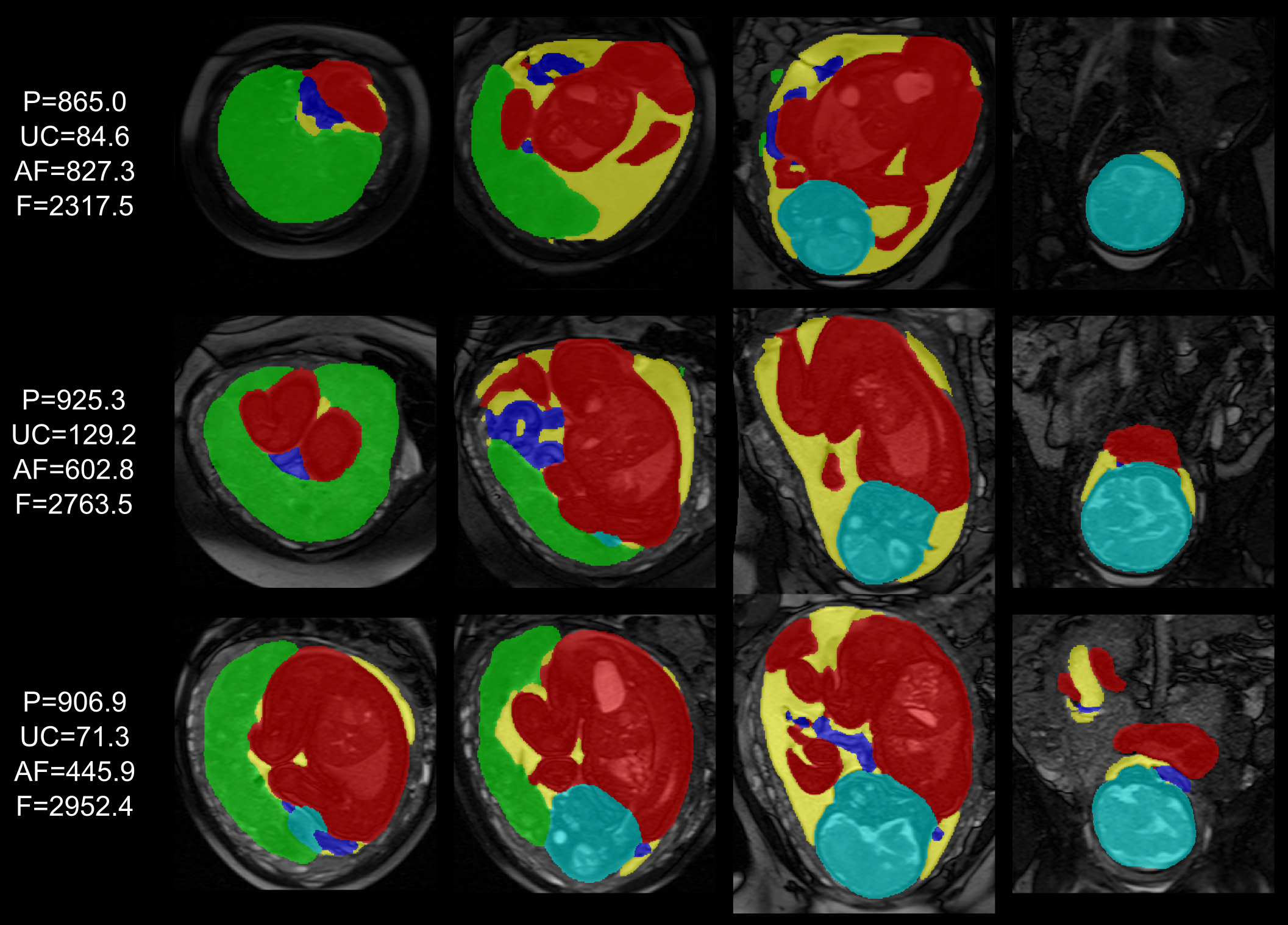}
    \caption{Retrospective testing: examples of predicted segmentations for three fetal subjects with different positions within the uterus. The calculated volume measurements (cc) of the key structures: placenta (P, green), umbilical cord (UC, dark blue), amniotic fluid (AF, yellow), and whole fetus (F, red and light blue).}
    \label{img:test-set-example}
\end{figure*}

Predicted segmentations for three randomly chosen exemplary fetal subjects from the retrospective testing cohort are shown in Fig.~\ref{img:test-set-example}. All 18 coronal stacks from 18 subjects were visually reviewed and deemed to be of acceptable quality. Quantitative results comparing network predictions to the average of manually refined segmentations (by two radiographers - R1 and R2 - and one researcher - R3) are summarised in Tab.~\ref{tab:volume_differences}. The model showed excellent performance on large ROIs, with high Dice scores and minimal volume discrepancies: $\approx 1\%$ for the fetus, $\approx 3\%$ for the placenta, and $\approx 2\%$ for the amniotic fluid. Performance was lower for the umbilical cord ($\approx 12\%$ average volume difference), due to its small size, variable shape and position, and limited visibility. Despite these challenges, the average Dice score for the cord remained high at 0.9.


\begin{table*}[ht!]
\centering
\resizebox{\textwidth}{!}{
\begin{tabular}{|l|c|c|c|c|} 
\hline
\rowcolor{gray!40} \textbf{} & \textbf{Fetus} & \textbf{Placenta} & \textbf{Umbilical Cord} & \textbf{Amniotic Fluid}  \\
\hline 
\textbf{Dice} & \multirow{3}{*}{\(0.99\pm0.01\)} & \multirow{3}{*}{\(0.98\pm0.02\)} & \multirow{3}{*}{\(0.91\pm0.08\)} & \multirow{3}{*}{\(0.99\pm0.01\)} \\
\textbf{(predictions vs. average R1-R3)} &  &  &  &   \\
\hline
\textbf{Absolute volume difference (cc)} & \multirow{3}{*}{\(8.57\pm8.88\)} & \multirow{3}{*}{\(19.34\pm25.74\)} & \multirow{3}{*}{\(10.10\pm11.23\)} & \multirow{3}{*}{\(7.03\pm8.61\)} \\
\textbf{(predictions vs. average R1-R3)} &  &  &  &   \\

\hline
\textbf{Relative volume difference (\%)} & \multirow{3}{*}{\(0.30\pm0.30\%\)} & \multirow{3}{*}{\(2.20\pm2.83\%\)} & \multirow{3}{*}{\(11.02\pm12.67\%\)} & \multirow{3}{*}{\(1.48\pm1.74\%\)} \\
\textbf{(predictions vs. average R1-R3)} &  &  &  &   \\
\hline
\end{tabular}
}
\caption{Retrospective testing: quantitative evaluation for 18 datasets. Comparison between the predicted segmentations from the nnUNet vs. the manually refined segmentations by three manual observers (average of all refinements).}

\label{tab:volume_differences}
\end{table*}

\subsection*{Normative ranges for late gestation datasets}
\begin{figure*}[h!]
    \centering
    \includegraphics[width=.85\textwidth]{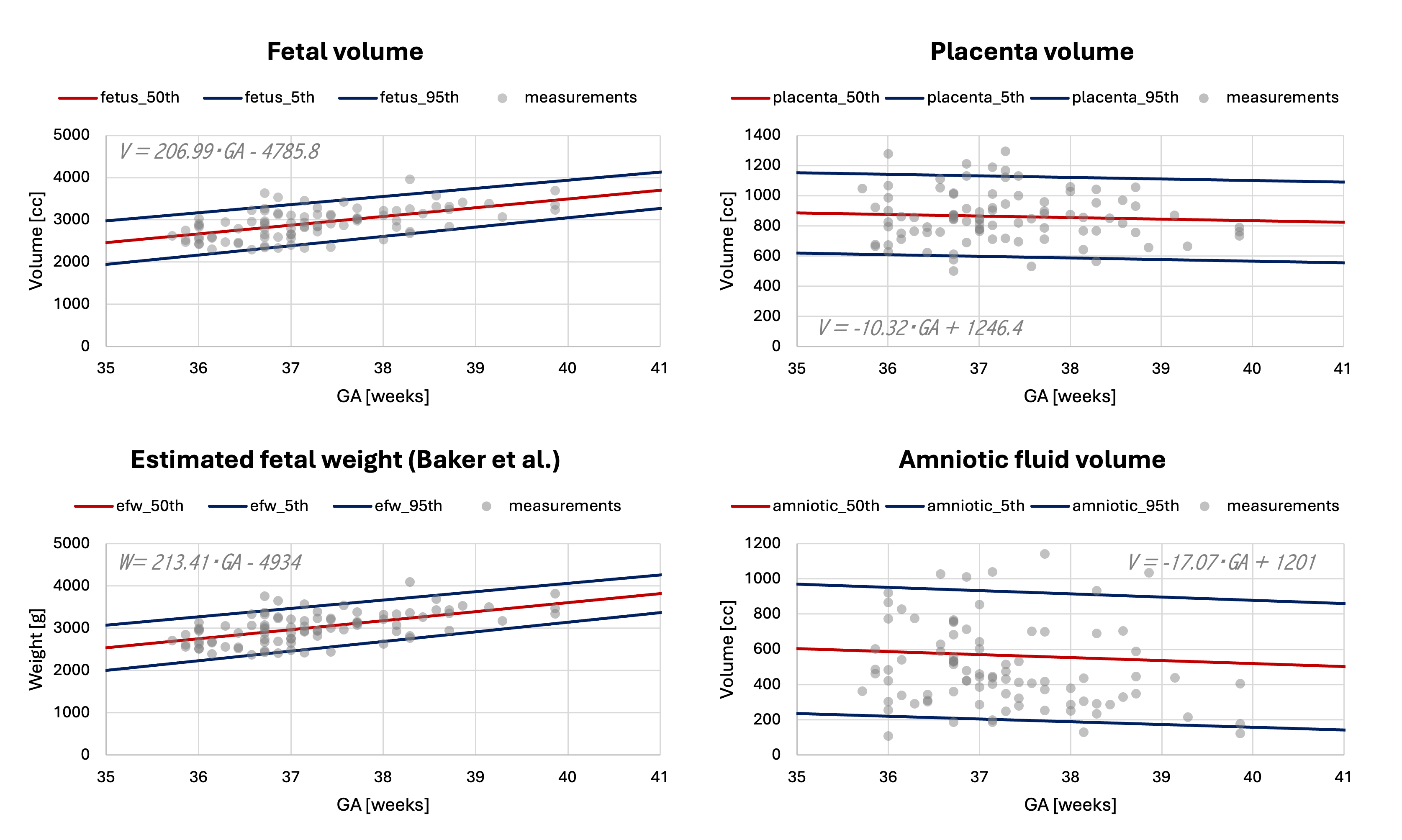}
    \caption{Normative ranges (50th, 5th and 95th centiles) for fetal, placenta and amniotic fluid volume and estimated fetal weight (Baker et al.), generated from 90 control subjects.}
    \label{fig:centiles}
\end{figure*}

Fig.~\ref{fig:centiles} presents the generated normative centiles for intrauterine regional volumetry and estimated fetal weight across the late GA range, based on the 90 control subjects (above mentioned) from the MiBirth study cohort. Segmentations required only minor refinements in fewer than 20\% of cases — primarily in challenging ROIs such as the umbilical cord, placenta, and fetal limbs, with minimal impact on the overall centile trends. As expected, fetal volume and estimated weight increased with gestational age, consistent with reported ranges \cite{SpecktorFadida2024}. Placental and amniotic fluid volumes exhibited greater variability in the later GA range, in line with findings from previous studies \cite{Peterson2022}. These results underscore the need for further research into the influence of maternal and fetal factors on normative modelling.

\begin{figure*}[h!]
    \centering
    \includegraphics[width=.95\textwidth]{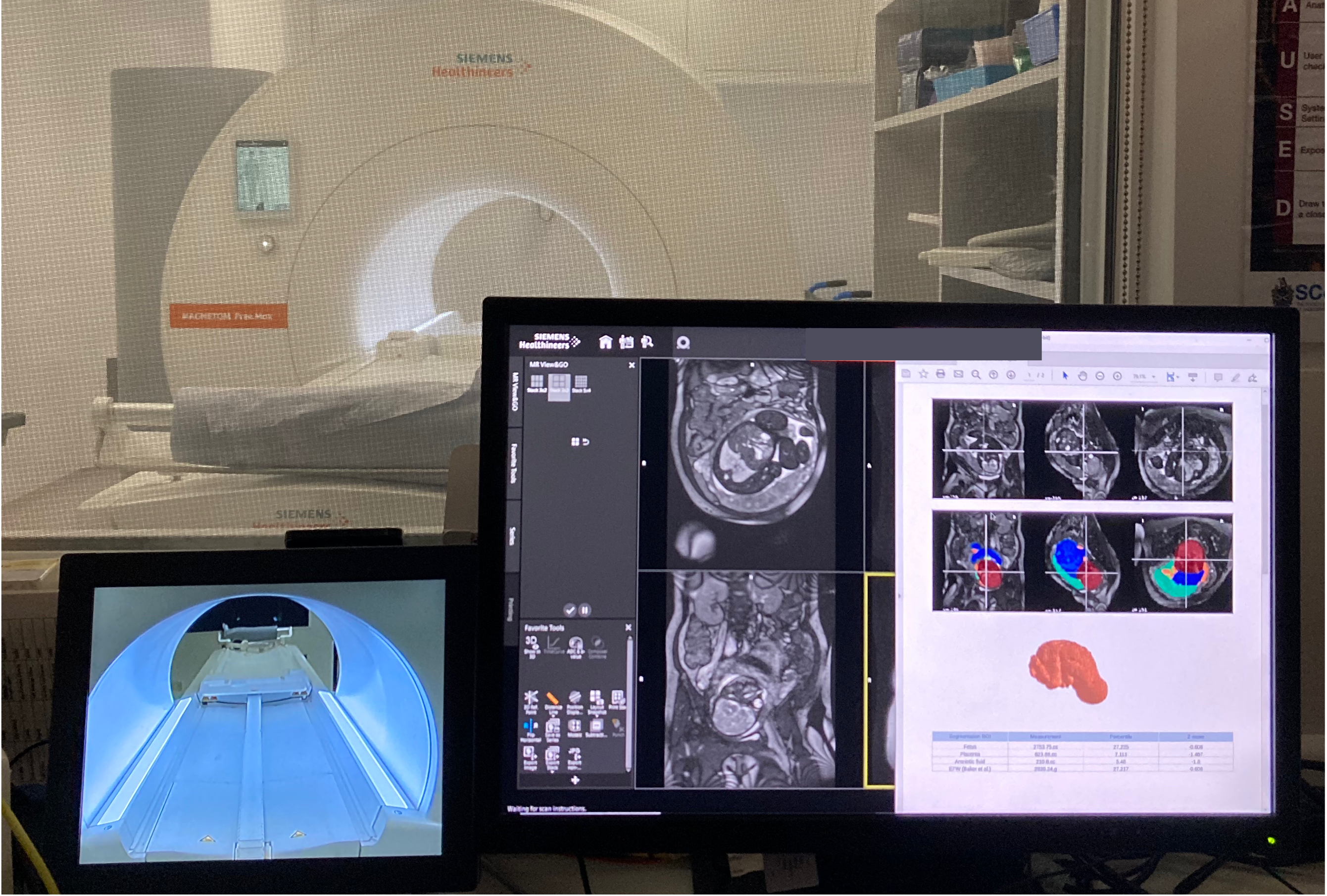}
    \caption{Automated volumetry report displayed on the scanner console immediately after the scan, shown with the coronal bSSFP acquisition used in the real-time volumetry pipeline on the MR viewer.}
    \label{fig:scanner-prospective}
\end{figure*}

\subsection*{Scanner-based automated reporting}

The complete segmentation and reporting pipeline, deployed on the 0.55T scanner using the FIRE framework, was prospectively evaluated in 50 fetal cases between 25–39 weeks GA. The segmentation network processed each stack in real-time, with an inference time of 45 seconds for multi-ROI label prediction. Volumetric segmentations were successfully generated in all cases, and structured reports were automatically produced and made available directly on the scanner workstation during the imaging session, enabling immediate review by clinical teams. Fig.~\ref{fig:scanner-prospective} shows an example of a report displayed on the scanner console.

\begin{figure*}[h!]
    \centering
    \includegraphics[width=.95\textwidth]{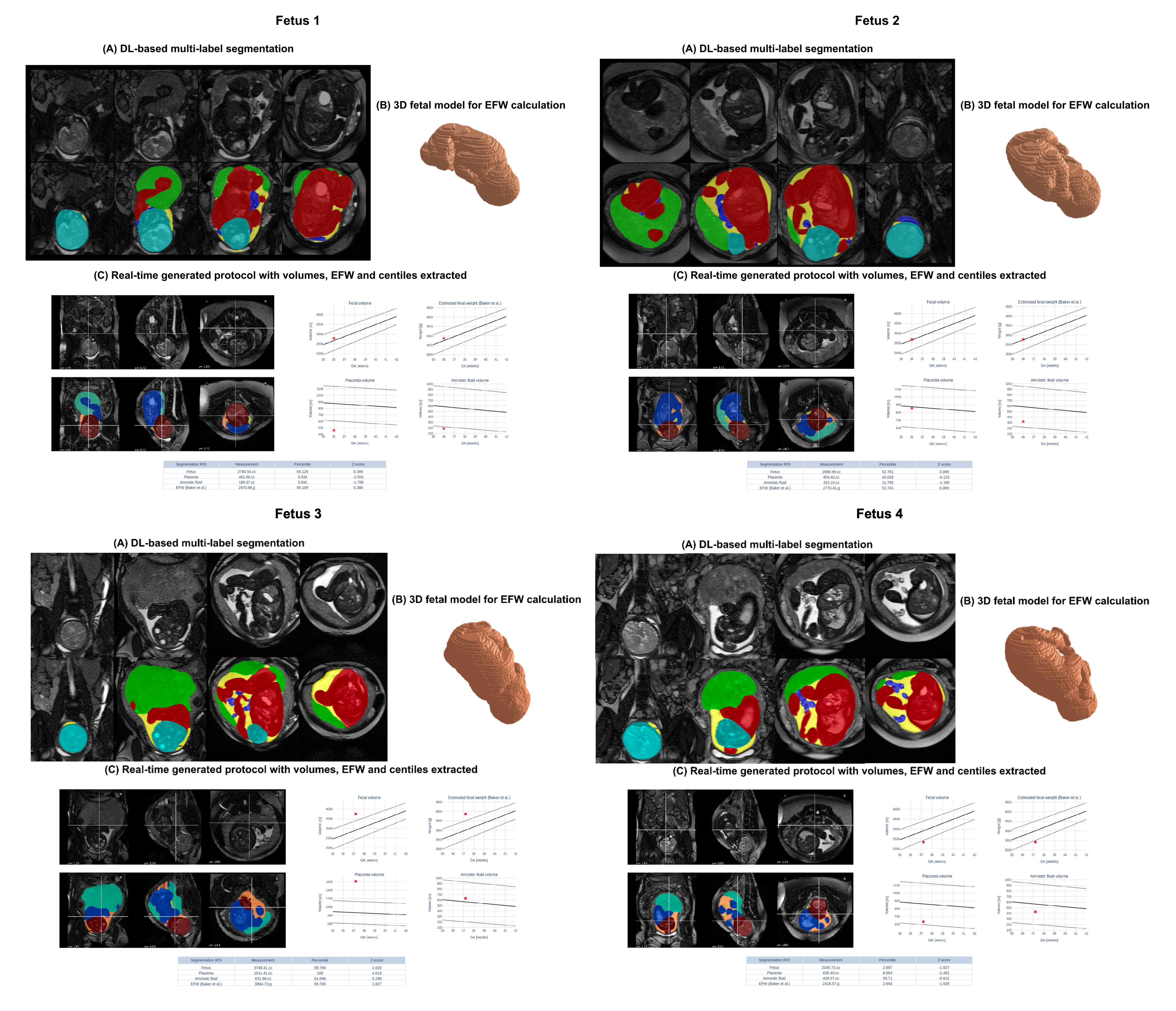}
    \caption{Prospective testing: examples of complete automated volumetry .pdf reports for four fetal subjects, depicting the acquired bSSFP images overlaid with the real-time segmentations (A), the 3D fetal model (B), and the real-time generated volumetry results (C).}
    \label{fig:examples-prospective}
\end{figure*}

\begin{figure*}[h!]
    \centering
    \includegraphics[width=.95\textwidth]{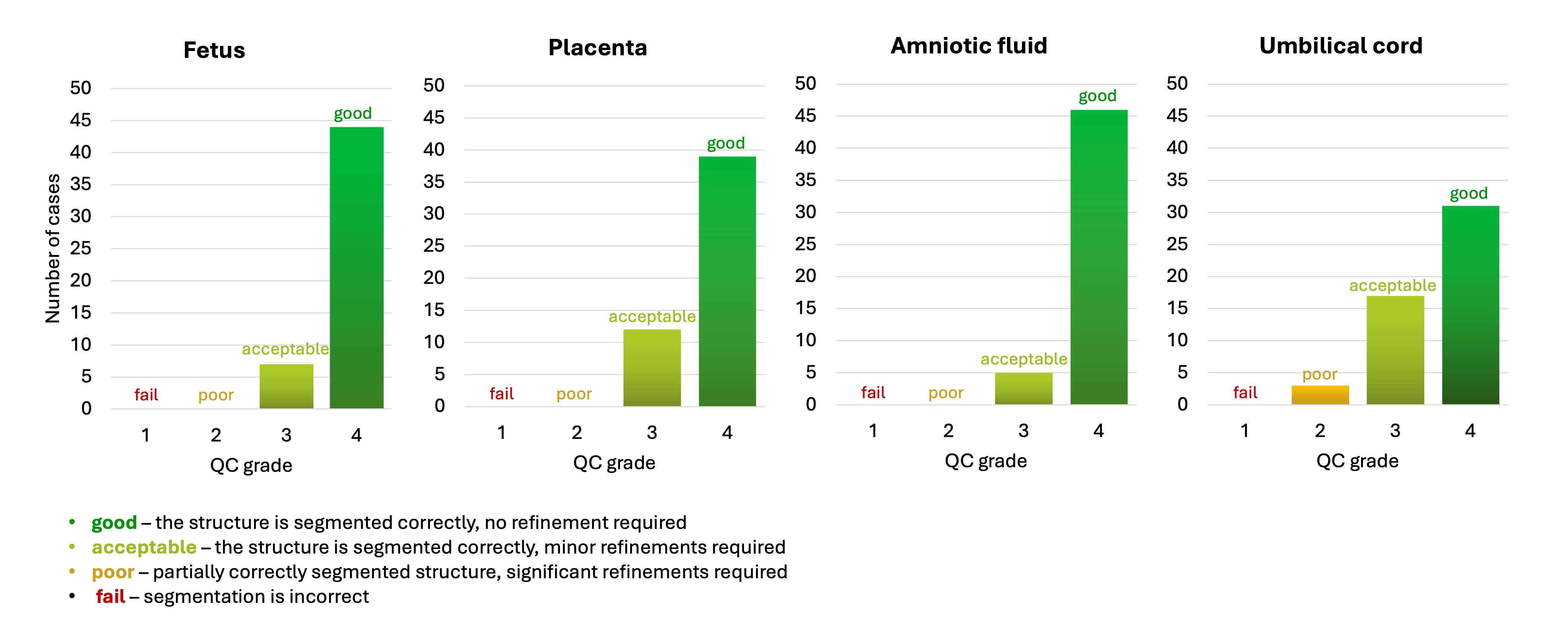}
    \caption{Prospective testing: qualitative evaluation results for 50 datasets. Grading of fetus, placenta, amniotic fluid and umbilical cord segmentation labels (fail, poor, acceptable, good).}
    \label{fig:prospective-qualitative}
\end{figure*}

Fig.~\ref{fig:examples-prospective} presents results from the full pipeline in four fetal cases, demonstrating the real-time extraction of key intrauterine structures and automated volumetry reporting. In several anecdotal cases, deviations from normative volumetric ranges correlated with ultrasound findings. For example, in Fetus 1, the placenta volume (461.7 cc) was below average for gestational age, and the amniotic fluid volume (186.4 cc) was at the lower end of normal. Ultrasound findings corroborated the low AFI, but no placenta anomalies were reported. In contrast, all volumetric measurements for Fetus 2 fell within expected ranges. Fetus 3 showed a total volume and estimated fetal weight (3748.4 cc) exceeding the 95th centile, while Fetus 4 measured below the 5th centile (2345.7 cc), aligning with the birthweight centiles for both of these cases: 97th and 7th, respectively.

Fig.~\ref{fig:prospective-qualitative} summarises qualitative scoring in 50 prospective cases. All segmentations were successfully generated using the integrated FIRE pipeline, with output labels and .pdf reports automatically saved to the scanner workstation. Over 95\% of segmented anatomical structures were rated as "excellent" or "good" in quality. The umbilical cord segmentations were characterised by lower scores due to minor errors caused by its inherent shape complexity and limited visibility. Importantly, no segmentation failures were recorded across any of the evaluated structures.

The clinical utility of the structured report was supported by verbal feedback from fetal MRI and obstetrics specialists (HW, SB, LS, MR). The added value of the comprehensive visualisation of all structures within the uterus, the inclusion of a 3D fetal model, and the integration of z-score-based quantification was particularly highlighted for the early identification of potential anomalies. Radiographers reported a positive user experience, emphasising the seamless integration of the pipeline into the existing MRI acquisition protocols. The tool was described as intuitive and easy to launch, requiring minimal changes to standard workflows.

\section*{Discussion}

This work introduces the first workflow for real-time, automated intrauterine volumetry reporting for fetal MRI, performed directly on the scanner during acquisition. The approach was integrated into the scanner environment (0.55T) via the FIRE framework and involves 1) an nnUNet network trained to automatically parcellate bSSFP whole uterus stacks into 5 regions: fetal body and head, placenta, amniotic fluid and umbilical cord; 2) a processing script that uses the segmentation results to subsequently generate a .pdf report that includes volumetric measurements, fetal weight estimation, and computed percentiles and z-scores against normative ranges, alongside visualisation of the segmented regions. The Gadgetron PC, connected to the scanner, launches the complete pipeline directly after stack acquisition. The resulting .pdf report is available for review directly on the scanner console during the scanning session.

Whilst deep learning segmentation of the whole fetus has already been addressed in several works (e.g., \cite{SpecktorFadida2024}), this is the first pipeline that combines segmentation of the fetus, placenta, umbilical cord and amniotic fluid in one network. It also includes the corresponding normative ranges specific to the late GA range and 0.55T fetal MRI. Joint volumetric analysis of these ROIs, in addition to fetal weight estimation, could potentially provide valuable information to radiologists and obstetricians to guide and support decision planning. Retrospective quantitative evaluation of the network demonstrated good performance for large structures, with less than 3\% relative volume differences after manual refinement. However, the umbilical cord was the most challenging for the model, due to suboptimal visibility, complex shape and high variability, leading to imperfect labelling in the training and testing datasets. These findings support the need for further training on a larger dataset to improve performance in anatomically complex structures.

Accurate volumetry of the fetus, placenta, amniotic fluid, and umbilical cord supports the identification of growth deviations (e.g. FGR or macrosomia), placental insufficiency, altered fluid dynamics, and potential umbilical cord-related risks. Multi-ROI segmentation also allows 3D rendering for visual assessment of the relative position information (e.g., umbilical cord vs. fetal neck, placenta location, etc.). When combined into an automated report with centiles and z-scores, this integrated analysis aids in stratifying risk, guiding delivery planning, and supporting clinical decision-making, particularly in complex or borderline cases.

The prospective evaluation of the scanner-deployed pipeline was successfully performed on 50 datasets, with the output reports and segmentations classified as acceptable/good for $>95\%$ of all ROIs in all cases. The feedback from both clinicians and radiographers was very positive, with clear added value of having estimated fetal weight and volumetry z-scores for identification of potential high-risk cases, as well as seamless integration into the scanning workflow. 

This confirms both the feasibility and utility of the proposed solution. One of the major advantages of scanner-integrated segmentation is the possibility to transfer the parcellation label files directly to the local archiving system as a part of the case files, without any additional processing on external workstations.

This prototype pipeline is the first step towards real-time scanner-based fetal MRI volumetric analysis. Automated reporting for fetal weight estimation and intrauterine regional volumetry could potentially reduce the radiologist's workload, thus paving the way for performing high-standard fetal reporting more widely in non-specialist centres.

\subsection*{Limitations and future work}

Although the segmentation network showed good performance on the test dataset, the main testing and training cohort was predominantly from the late GA range, singleton pregnancy, normal fetal anatomy, and from the same 0.55T acquisition protocol. Further retraining of the model on early GA datasets, fetal structural anomalies and higher-field strength protocols is required for translation to clinical practice. This will additionally require the generation of normative growth models for the whole duration of the second and third trimesters. 

In addition, integration of automated image quality control \cite{Sanchez2024} and image restoration via correction of motion and intensity artefacts (e.g., shading and banding) \cite{Lim2023} as a preprocessing step can further improve the accuracy of segmentations. For extreme motion corruption, detection and reacquisition of bSSFP stacks might also be required \cite{AvilesVerdera2025}. 

Further sub-parcellation of the fetal ROI into brain tissue regions \cite{Uus2023Bounti} and body organs \cite{Uus2024bodyorgans} could potentially enable full volumetric reporting from a single stack. Segmentation of additional maternal anatomy structures (myometrium and cervix), as well as optimisation of the pipeline for multiple gestation pregnancies, could also be beneficial. The report structure will be extended with 3D visualisation of all structures and relative position assessment.

\section*{Conclusion}

We presented the first scanner-deployed automated fetus, placenta, amniotic fluid and umbilical cord volumetry reporting pipeline for fetal MRI. It enables fast DL segmentation of the whole uterus directly from 0.55T bSSFP stacks during acquisition, with a .pdf report with z-scores and visualisation of the results available on the scanner console. The pipeline was successfully evaluated both retrospectively (18 cases) and prospectively (50 cases), and demonstrated good segmentation quality and positive user feedback from clinicians and radiographers. Real-time fetal weight estimation and multi-ROI volumetry reporting have the potential to provide valuable information to radiologists, thus opening the possibility for high-standard fetal MRI assessment in non-specialist centres. Future work will focus on extending the GA range, optimising the pipeline for 1.5T and 3T, and further sub-parcellating the fetal brain and body ROIs. We also plan to integrate image restoration and quality control measures.


\section*{Acknowledgements}
We thank everyone involved in the acquisition and analysis of the datasets at the Department of Early Life Imaging at King's College London and St Thomas' Hospital. The authors thank all participants and their families. 

This work was supported by the MRC grant [MR/X010007/1], the Wellcome Trust, Sir Henry Wellcome Fellowship to Jana Hutter [201374/Z/16/Z], {DFG Heisenberg funding [502024488]}, the UKRI, FLF to Jana Hutter [MR/T018119/1], NIHR Advanced Fellowship awarded to Lisa Story [NIHR30166], the Wellcome Trust and EPSRC IEH award [102431] for the iFIND project [WT 220160/Z/20/Z], the Wellcome/ EPSRC Centre for Medical Engineering at King’s College London [WT 203148/Z/16/Z], the NIHR Clinical Research Facility (CRF) at Guy’s and St Thomas’ and by the National Institute for Health Research Biomedical Research Centre based at Guy’s and St Thomas’ NHS Foundation Trust and King’s College London.

The views expressed are those of the authors and not necessarily those of the NHS, the NIHR or the Department of Health.

\section*{Author contributions}
S.N.S. and A.U. contributed equally to this work and the manuscript. S.N.S. trained the deep learning model, and S.N.S. and J.A.V. developed and tested the pipeline for integration into the scanner environment. A.U. formalised the segmentation protocol, prepared training datasets, and developed an automated reporting script. H.W. contributed to analysis and acquisition of the datasets. J.A.V. and J.H contributed to optimisation of the acquisition protocol and the acquisition of the datasets. K.S.C., W.N., S.B contributed to preparation of the training dataset, acquisition of the datasets, and the testing of the pipeline. D.C., T.W., M.v.P., J.K.S., J.M., K.P., D.L contributed to preparation of the training dataset derived from other projects. V.K. contributed to acquisition and analysis of the datasets. D.S. and A.D. supervised various components of the project. J.V.H. and L.S. provided datasets and supervised various components of the project. J.H. and M.R. provided datasets and supervised the project. All authors reviewed the manuscript.





\end{document}